\documentclass[runningheads]{cl2emult}

\usepackage{makeidx}  
\usepackage{graphicx} 
\usepackage{subeqnar} 
\usepackage{multicol} 
\usepackage{cropmark} 
\usepackage{eso}      
\makeindex            

\begin{document}
\title*{A scenario for the formation of dwarf galaxies
with an anomally low dark-matter content}
\toctitle{A scenario for the formation of dwarf galaxies
\protect\newline with an anomally low dark-matter content}
%
%
\titlerunning{A scenario for the formation of dwarf galaxies}
%
\author{Peter Berczik
\and Sergei G. Kravchuk}
\authorrunning{Peter Berczik and Sergei G. Kravchuk}
%
%
\institute{Main Astronomical Observatory of
           Ukrainian National Academy of Sciences \\
           UA-03680, Golosiiv, Kiev-127, Ukraine.
           E-mail: {\tt berczik@mao.kiev.ua}}


\maketitle             


\begin{abstract}

The formation and evolution of a low mass galaxy in the gravitational 
field of a massive disk galaxy (like the Milky Way) has been studied. 
Numerical simulations of complex gas-dynamic flows are based on our own 
variant of the Chemo--Dynamical Smoothed Particle Hydrodynamical ({\bf 
CD--SPH}) approach, which incorporates star formation. The dynamics of 
the dark matter was treated as a standard N--body problem. It is shown 
that the satellite galaxy effectively looses its dark matter component 
due to the strong tidal influence of the massive galaxy, while the gas, 
owing to its strong dissipative nature, forms an almost 
dark-matter--free dwarf galaxy.

\end{abstract}


\section{Method}

Dwarf galaxies are the most dominant population of the present--day
Universe \cite{M98}. In the standard cold dark matter model low--mass
galaxies are considered to be the first luminous objects. If the present
dwarfs are dark matter dominated, they could account for a significant
fraction of the mass of galaxy clusters. Besides their numerosity,
dwarfs seem to be the simplest galactic system available for reliable
observational study and theoretical modelling.

The hydrodynamical simulations are based on our own variant of the 
Chemo--Dynamical Smoothed Particle Hydrodynamics ({\bf CD--SPH}) 
approach, including feedback through star formation phenomena. The 
dynamics of the dark matter component are treated within a standard 
N--body approach. Thus, the satellite galaxy consists of dark matter, 
gas and "star" particles. For a detailed description of the {\bf 
CD--SPH} code (the star formation algorithm, the SNII, SNIa and PN 
production, and the chemical enrichment) the reader is referred to 
\cite{BerK96,Ber99}. It is to be noted that the code was slightly 
modified for the present problem so as to include the photometric 
evolution of each "star" particle, based on the idea of the Single 
Stellar Population (SSP) \cite{BCF94,TCBF96}. At each time-step, 
absolute magnitudes: M$_U$, M$_B$, M$_V$, M$_R$, M$_I$, M$_K$, M$_M$ and 
M$_{bol}$ are defined separately for each "star" particle. The SSP 
integrated colours (UBVRIKM) are taken from \cite{TCBF96} (Table 2). The 
spectro--photometric evolution of the overall ensemble of the "star" 
particles forms the Spectral Energy Distribution (SED) of the satellite.


\section{Model parameters}

The central host galaxy is assumed to be a Milky-Way--like system. It is
modelled as a Disk + Bulge + Halo system according to \cite{DC95}.

The initial satellite proto-galaxy is taken to be a gas--rich and dark
matter dominated proto--dwarf with a total mass $M_{SAT} = 2 \cdot 10^9
\; M_\odot $ and an initial radius $ R_{SAT} = 5 \; {\rm kpc} $. The
baryonic (i.e. initial "gas") mass of the proto--dwarf is $
M^{gas}_{SAT} = 2 \cdot 10^8 \; M_\odot $, and the initial Dark-Matter
(DM) mass is $ M^{dm}_{SAT} = 1.8 \cdot 10^9 \; M_\odot $. Initially the
centre of the proto--dwarf is located at Galactocentric coordinates (0,
0, 150 kpc). The initial velocity field of this object is defined as the
sum of the orbital motion in the host galaxy in the polar plane (YZ) and
the rotation around its own rotational axis (parallel to the X axis of
host galaxy). The initial velocity of orbital motion is: $ V_y = 75 \;
{\rm km/sec} $. The dimensionless angular velocity vector of the
rotation is $ {\bf \Omega} = (1, 0, 0) $.  As the system unit for this
physical parameter we use the value: $ \Omega_0 = \sqrt{G \cdot
M_{SAT}/R_{SAT}} / R_{SAT} $.


\section{Results}

Extensive numerical simulation show that model evolution results in the 
formation of a baryon-dominated satellite galaxy. The basic process 
keeping this scenario is the dissipative nature of the baryonic 
component of the proto--dwarf, which easily forms a compact structure 
with a radius smaller than the tidal radius. The practically 
dissipationless dark-matter component forms an extended structure which 
is effectively destroyed via tidal influence of the massive galaxy. This 
mechanism is very interesting in the context of the possible existence 
of of dSph galaxies with a low dark-matter content.


{\bf Acknowledgements} The authors are grateful to Pavel Kroupa for
stimulating this work and L.S. Pilyugin and Yu.I. Izotov for fruitful
discussions during the process of preparing this work. The work was
partially supported by NATO grant {\bf NIG 974675}.


\clearpage
\addcontentsline{toc}{section}{Index}
\flushbottom
\printindex

\end{document}